\documentclass{article}
\pdfoutput=1
\usepackage{arxiv}
\usepackage{algorithmic}
\usepackage{textcomp}
\def\BibTeX{{\rm B\kern-.05em{\sc i\kern-.025em b}\kern-.08em
    T\kern-.1667em\lower.7ex\hbox{E}\kern-.125emX}}

\usepackage[dvips]{graphicx}
\usepackage{cite,amssymb,amsmath,amsfonts,epsfig,color,epstopdf}
\newcommand{\beq}{\begin{equation}}
\newcommand{\eeq}{\end{equation}}
\newcommand{\beqn}{\begin{eqnarray}}
\newcommand{\eeqn}{\end{eqnarray}}
\DeclareMathOperator*{\argmin}{arg\,min}
\def\bmath#1{\mbox{\boldmath$#1$}}

\usepackage{algorithm,algorithmic}
\floatname{algorithm}{Algorithm}

\usepackage{natbib}

\title{Hint assisted reinforcement learning: an application in radio astronomy}

\author{ Sarod Yatawatta\\
ASTRON, The Netherlands Institute for Radio Astronomy,\\Dwingeloo, The Netherlands\\ yatawatta@astron.nl } 

\begin{document}
\maketitle
\begin{abstract}
  Model based reinforcement learning has proven to be more sample efficient than model free methods. On the other hand, the construction of a dynamics model in model based reinforcement learning has increased complexity. Data processing tasks in radio astronomy are such situations where the original problem which is being solved by reinforcement learning itself is the creation of a model. Fortunately, many methods based on heuristics or signal processing do exist to perform the same tasks and we can leverage them to propose the best action to take, or in other words, to provide a `hint'. We propose to use `hints' generated by the environment as an aid to the reinforcement learning process mitigating the complexity of model construction. We modify the soft actor critic algorithm to use hints and use the alternating direction method of multipliers algorithm with inequality constraints to train the agent. Results in several environments show that we get the increased sample efficiency by using hints as compared to model free methods.
\end{abstract}

\section{Introduction}

Deep reinforcement learning (RL) is moving beyond its breakthrough mainstream applications \citep{Atari,silver2016mastering,Levine2015} into diverse and fringe disciplines, for example into radio astronomy \citep{Y2021}. Many of these applications of RL use model free methods for training the RL agent, where the agent interacts with the environment via the reward for each action being taken. Model based reinforcement learning \citep{Wang2019, janner2019trust, Clavera2018, Clavera2020} builds an additional internal model of the environment to predict next states given the current state and action being taken. Therefore model based RL is able to mimic the environment thus requiring fewer interactions with the environment. Hence model based RL is generally more efficient in terms of the data needed for learning.

Modern radio astronomy is heavily data intensive, and new instruments are being built that will generate petabytes of stored data at terabits per second rates. We give a brief and simplified introduction to radio interferometry here: Celestial signals are sampled by an array of sensors on Earth and sent to a correlator. Signals from each sensor is correlated with the signals from other sensors to form correlations of interferometric pairs. Finally, an image of the celestial sky is formed by performing a Fourier transform on the sphere using the correlated data. The transformation of the raw data produced by the sensor array to an image of the celestial sky requires many precise operations on the data. Machine learning has been applied to almost every operation in this whole transformation, for example in data inspection \citep{Mesarcik}, in interference mitigation \citep{Sadr2020} and in image formation \citep{WuLiu2022}. Calibration is another major operation performed on the data that builds a model for the systematic errors affecting the data, enabling correction for such errors. Reinforcement learning has already been applied to fine tune calibration \citep{Y2021}. 

In this paper, we focus on improving the data efficiency in training an RL agent to perform tasks in fine tuning radio interferometric calibration. Calibration itself is a task that builds a model and the direct application of model based RL to learn this task is infeasible. This is mainly due to the complexity of creating two models as well as ill-conditioning. As an alternative, we modify the agent to use a hint for the possible action to take given the state information. This hint can be generated by any means, for example using signal processing or by heuristics or even by a human expert. This hint does not have to be perfect as well. We can use any metric to measure the distance between the action produced by the agent and the hint and we keep this distance below a threshold. We modify the soft actor-critic (SAC) algorithm \citep{SAC,SAC1} with this concept of using hints and that boils down to policy optimization under an inequality constraint. We use the alternating direction method of multipliers (ADMM) \citep{boyd2011} algorithm with inequality constraints \citep{Giesen} to train the actor.

\paragraph{Related work}
\begin{itemize}
  \item The use of RL in data processing tasks such as calibration is not uncommon. For example, \citep{Ich2021} use RL to accelerate the solving of quadratic programs by fine tuning hyperparameters. Both \citep{Y2021} and \citep{Ich2021} are using model free RL and therefore would benefit from model based RL and using hints as proposed by this work.
  \item The use of ADMM in policy optimization has been done before, for example \citep{mordatch2014combining} use ADMM in robot trajectory optimization. Similarly, \citep{Levine2015} use Bregman divergence based ADMM for learning visuomotor policies. Both these methods have used equality constraints and our work uses inequality constraints with ADMM \citep{Giesen} enabling flexible constraints, especially when the hints are not entirely accurate.
  \item Model based RL combined with a model free learner is used by \citep{Anusha2017} in a model predictive control setting. The model free learner is initialized by a dynamics model. The update of the dynamics model and the model free learner is done alternatively. Our work is an improvement of \citep{Anusha2017}, especially when the dynamics model has high complexity, by replacing the dynamics model by hints and by using inequality constraints to allow for some inaccuracies in the provided hints. In other words, our method is simpler than the method of \citep{Anusha2017} provided that there is a way to generate hints that are accurate enough.
  \item Constrained optimization has been used in RL, e.g., \citep{farahmand2008regularized,yang2021wcsac,Zhou2022}, mainly in safety critical applications or to limit a use of a resource such as fuel. Such work apply penalties to the reward or constrain the critic during training. In contrast, our work applies constraints on the policy or the actor.
\end{itemize}

The rest of the paper is organized as follows: In section \ref{sac}, we describe the SAC algorithm using hints. In section \ref{cal}, we provide an overview of radio interferometric calibration and how RL is being used in calibration. Next, in section \ref{results}, we provide results based on several environments and finally, we draw our conclusions in section \ref{conc}.

{\em Notation}: Lower case bold letters refer to column vectors (e.g. ${\bmath s}$). Upper case bold letters refer to matrices (e.g. ${\bf { C}}$). The matrix inverse, transpose, Hermitian transpose, and conjugation are referred to as $(.)^{-1}$, $(.)^{T}$, $(.)^{H}$, $(.)^{\star}$, respectively. The matrix Kronecker product is given by $\otimes$. The vectorized representation of a matrix is given by $\mathrm{vec}(.)$. The identity matrix of size $N\times N$ is given by ${\bf {I}}_N$.  Estimated parameters are denoted by a hat, $\widehat{(.)}$. All logarithms are to the base $e$, unless stated otherwise.

\section{Soft actor critic with hints\label{sac}}
In this section, we describe the SAC algorithm \citep{SAC1} that has been modified to use hints. Similar modifications can be done to other RL algorithms such as TD3 \citep{TD3}. At step index $t$, ${\bf s}_t \in \mathcal{S}$ is the state vector and ${\bf a}_t \in \mathcal{A}$ is the action being taken, which leads to the next state ${\bf s}_{t+1}$ giving us the reward $r({\bf s}_t,{\bf a}_t) \in \mathbb{R}$. The critic is represented by $Q_{\theta}({\bf s}_t,{\bf a}_t)$ parameterized by ${\theta}$ and the policy is given by $\pi_{\phi}({\bf a}_t|{\bf s}_t)$ which is parameterized by ${\phi}$.
For off policy learning, the past experience is collected in the replay buffer $\mathcal{D}$ that collects tuples $\left({\bf s}_t,{\bf a}_t,r({\bf s}_t,{\bf a}_t),{\bf s}_{t+1}\right)$. An offline (or target) critic is parameterized by $\overline{\theta}$ that is updated at delayed intervals with weight $\tau$. 

The value function is given by
\beq \label{value}
V_{\overline{\theta}}({\bf s}_t) = E_{{\bf a} \sim \pi_\phi} \left[Q_{\overline{\theta}}({\bf s}_t,{\bf a}_t) - \alpha \log \pi_{\phi}({\bf a}_t|{\bf s}_t) \right]
\eeq
and the critic is updated by minimizing the loss
\beqn \label{qloss}
\lefteqn{ J_Q(\theta) 
 =E_{({\bf s}_t,{\bf a}_t) \sim \mathcal{D}} } \\\nonumber 
&& \left[ \frac{1}{2} \left( Q_{\theta} ({\bf s}_t,{\bf a}_t)  - \left( r({\bf s}_t,{\bf a}_t) + \gamma E_{{\bf s}_{t+1} \sim \mathrm{p}} \left[ V_{\overline{\theta}}({\bf s}_{t+1}) \right] \right) \right)^2 \right]
\eeqn
where $\mathrm{p}$ is the state transition probability from ${\bf s}_t$  to ${\bf s}_{t+1}$ due to ${\bf a}_t$.

We introduce the hint ${\bf h}_t \in \mathcal{A}$ as follows. We introduce a constraint $c({\bf a}_t,{\bf h}_t) < \delta$ where $c(\cdot,\cdot)$ can be any metric, such as the mean squared error (MSE) or Kullback Liebler divergence (KLD). The distance measured by $c(\cdot,\cdot)$ is kept below the threshold $\delta \in \mathbb{R^+}$. The hint can be made stochastic as well, for example by providing both the mean and the covariance for ${\bf h}_t$ but in this work we only consider deterministic hints.

For the policy update, we need to incorporate the aforementioned constraint. Following \citep{Giesen}, we build 
\beq
g({\bf a}_t,{\bf h}_t) \buildrel\triangle\over = \left(\left[c({\bf a}_t,{\bf h}_t)-\delta\right]_{+}\right)^2
\eeq
where $[x]_{+}=x$ if $x>0$ and otherwise $[x]_{+}=0$.
The augmented Lagrangian is given by
\beqn \label{aug}
\lefteqn{\mathcal{L}_{\pi}(\phi)}\\\nonumber
&&=E_{{\bf s}_t\sim \mathcal{D}, {\bf a}_t\sim \pi_{\phi}}\left[ \alpha \log \left( \pi_{\phi}({\bf a}_t|{\bf s}_t) \right) - Q_{\theta}({\bf s}_t,{\bf a}_t)\right]\\\nonumber
&&+ \frac{\rho}{2}E_{{\bf a}_t\sim \pi_{\phi}}\left[g({\bf a}_t,{\bf h}_t)^2\right] + \mu E_{{\bf a}_t\sim \pi_{\phi}}\left[g({\bf a}_t,{\bf h}_t) \right]
\eeqn
where $\rho \in \mathbb{R^+}$ is the regularization parameter and $\mu \in \mathbb{R}$ is the Lagrange multiplier. Using (\ref{aug}), we can apply ADMM \citep{Giesen} for policy update and at each iteration, we update the Lagrange multiplier as
\beq
\mu \leftarrow \mu + \rho E_{{\bf a}_t\sim \pi_{\phi}}\left[g({\bf a}_t,{\bf h}_t) \right].
\eeq

The complete SAC with hints is given in algorithm \ref{algSAC}.
\begin{algorithm}
\caption{Soft actor critic with hints}
\label{algSAC}
\begin{algorithmic}[1]
  \REQUIRE $\theta_1$,$\theta_2$,$\phi$, $C$: cadence of Lagrange multiplier update, $\rho$: regularization factor, $\alpha$: temperature, $\gamma$: discount factor, $\tau$: weight update factor, $\lambda$: learning rate
  \STATE Initialize $\overline{\theta}_i \leftarrow \theta_i$ for $i\in[1,2]$, $\mu \leftarrow 0$, $\mathcal{D}\leftarrow$ \{take random steps to fill $D$ tuples\}
\FOR{each iteration}
  \FOR{each environment step}
    \STATE ${\bf a}_t \sim \pi_{\phi}({\bf a}_t|{\bf s}_t)$
    \STATE Get ${\bf s}_{t+1}$ and $r({\bf s}_t,{\bf a}_t)$ from the environment
    \STATE $\mathcal{D}\leftarrow \mathcal{D}\cup({\bf s}_{t},{\bf a}_t,r({\bf s}_t,{\bf a}_t),{\bf s}_{t+1})$
  \ENDFOR
  \FOR{each learning step}
  \STATE $\theta_i \leftarrow \theta_i - \lambda \nabla_\theta J_Q(\theta)$ for $i \in [1,2]$
  \STATE $\phi \leftarrow \phi - \lambda \nabla_\phi \mathcal{L}_{\pi}(\phi)$
  \STATE $\overline{\theta}_i \leftarrow \tau \theta_i +(1-\tau) \overline{\theta}_i$ for $i \in [1,2]$
  \IF{learning step is a multiple of $C$}
    \STATE $\mu \leftarrow \mu + \rho E_{{\bf a}_t\sim \pi_{\phi}}\left[g({\bf a}_t,{\bf h}_t)\right]$
  \ENDIF
  \ENDFOR
\ENDFOR
\end{algorithmic}
\end{algorithm}
Similar to \citep{SAC1}, the practical implementation of algorithm \ref{algSAC} use two target critic networks and use the minimum Q-value of the two for evaluating (\ref{value}).
\section{Radio interferometric calibration\label{cal}}
\begin{figure}[h]
\begin{minipage}{0.98\linewidth}
\begin{center}
\epsfig{figure=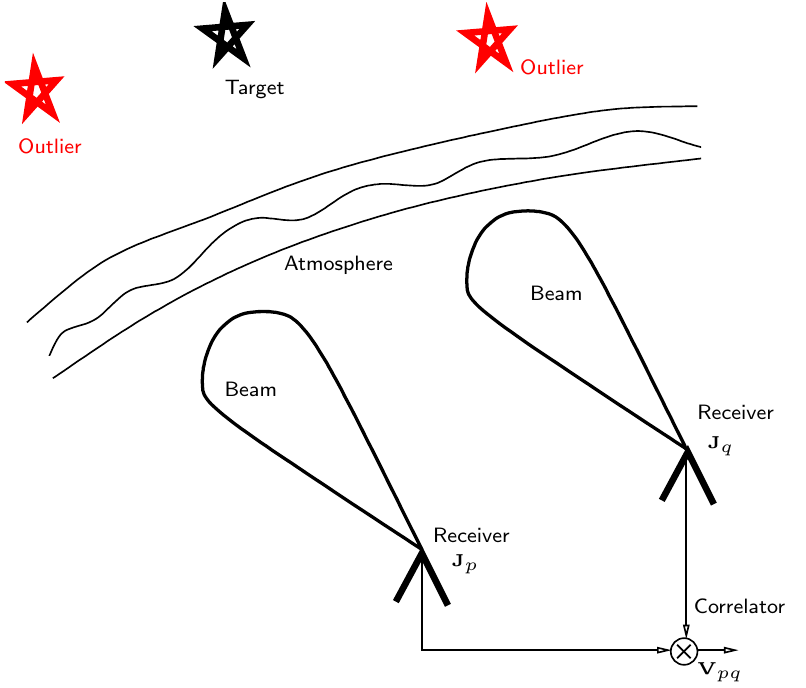,width=8.0cm}\\
\end{center}
\end{minipage}
  \caption{A simple interferometer observing a target direction while there are two outlier sources appearing as well.\label{interf}}
\end{figure}

In this section, we give a brief overview of radio interferometric calibration and the problem pertaining to calibration that is being solved by RL. In Fig. \ref{interf}, we show a simple schematic of a pair of receivers on Earth collecting data from the sky and correlating that data, forming an interferometer.

Given a pair of receivers $p$ and $q$ on Earth, the output at the correlator can be given as \citep{HBS}
\beq \label{V}
{\bf V}_{pq}=\sum_{i=0}^K {\bf J}_{pi} {\bf C}_{pqi} {\bf J}_{qi}^H + {\bf N}_{pq}
\eeq
where ${\bf V}_{pq}$ ($\in {\mathbb C}^{2\times 2}$) is a $2\times 2$ matrix of complex numbers that are the observed data. This data are assumed to be composed of a signal corresponding to a source in the sky being observed (target) and $K$ signals of outlier sources that act as interference. The signal from each source in the sky at the receiver is corrupted by systematic errors (varying both in time and in frequency) that are represented as ${\bf J}_{pi}$,${\bf J}_{qi}$ ($\in {\mathbb C}^{2\times 2}$) in (\ref{V}). The uncorrupted signal of each source is given by ${\bf C}_{pqi}$ ($\in {\mathbb C}^{2\times 2}$) and this is fairly stable and can be pre-computed. Additionally, we also have a contribution from noise, which is modeled by ${\bf N}_{pq}$ ($\in {\mathbb C}^{2\times 2}$).

Given $N$ receivers, we can form $N(N-1)/2$ interferometers and by observing over a long time period and a large bandwidth, we can increase the number of collected data points. Calibration in a nutshell is finding ${\bf J}_{pi}$,${\bf J}_{qi}$ in (\ref{V}), given ${\bf V}_{pq}$ and ${\bf C}_{pqi}$ for all $p$,$q$ and $i$. There are specialized software already to perform calibration, see for example \citep{InPar}. Our interest in using RL is to select the directions to calibrate, in other words, given $K$ possible directions in (\ref{V}) $i\in[1,K]$, select the subset of $i$ (say $\mathcal{I}$) to get the best output data quality with minimum computational cost spent in calibration. Note that the target direction corresponds to $i=0$ and is always included in the calibration. We should also mention here that calibration is computationally demanding, and it has to be performed for each data point of many data points covering a large time and frequency interval, accumulating into thousands of separate calibrations.

We use an approach based on the Akaike information criterion (AIC) \citep{Akaike} to generate the hint by selecting the best possible subset of $i \in[1,K]$ (or $\mathcal{I}$) to use in calibration. We rewrite (\ref{V}) in vector form as
\beq \label{vpq}
{\bf v}_{pq}=\sum_{i=0}^K {\bf s}_{pqi} +{\bf n}_{pq}
\eeq
where ${\bf s}_{pqi}=({\bf J}_{qi}^{\star}\otimes{\bf J}_{pi}) \mathrm{vec}({\bf C}_{pqi})$, ${\bf v}_{pq}=\mathrm{vec}({\bf V}_{pq})$, and ${\bf n}_{pq}=\mathrm{vec}({\bf N}_{pq})$. We can stack up ${\bf v}_{pq}$ in (\ref{vpq}) for all possible $p,q$ and for all time and frequency ranges within which a single solution for calibration is obtained. Let us call this ${\bf y}$. We can do the same for the right hand side of (\ref{vpq}) for each direction $i$ to get ${\bf s}_i$ and also for the noise to get ${\bf n}$. Let us rewrite (\ref{vpq}) after stacking as
\beq
{\bf y}={\bf s}_0+\sum_{i\in \mathcal{I}} {\bf s}_{i} +{\bf n} 
\eeq
where we have ${\bf s}_0$ for the model of the target direction and the remaining directions are taken from the set $\mathcal{I}$. For this setup, after calibration, we find the residual signal as
\beq \label{res}
{\bf r}={\bf y}-\widehat{\bf s}_0-\sum_{i\in \mathcal{I}} \widehat{\bf s}_{i}
\eeq
where $\widehat{\bf s}_i$ is the model constructed by calibration (so not necessarily the true model). Using (\ref{res}), we find the AIC as
\beq \label{AIC}
AIC_{\mathcal{I}}=\left(\frac{N \sigma_{\bf r}}{\sigma_{\bf y}}\right)^2+N|\mathcal{I}|
\eeq
where $\sigma_{\bf y}$ and $\sigma_{\bf r}$ are the standard deviations of ${\bf y}$ and ${\bf r}$ in (\ref{res}), respectively. The cardinality of $\mathcal{I}$ is given by $|\mathcal{I}|$.

With $K$ directions, we have $2^K$ possible configurations for $\mathcal{I}$. By evaluating (\ref{AIC}) for each of them, it is possible to select $\mathcal{I}$ that minimizes (\ref{AIC}), which is what is done in current practice. There are several shortcomings of this approach. First, $2^K$ is a large number, especially when we need to perform a computationally expensive calibration for each one of them (in real-time if needed). Secondly, we ignore the underlying global dependencies between different observations (i.e., when the target is in different directions in the sky). The celestial sky is very stable in terms of source directions and their intensities. However, because we evaluate (\ref{AIC}) for each observation individually, there is no way of incorporating this stable behavior with respect to multiple observations into one. This is the motivation for using RL for the determination of $\mathcal{I}$ given any observation. We intend to cut down the number of calibration runs needed to evaluate (\ref{AIC}) from $2^K$ to a lower value around $K$.

Let us explain the mapping between $j \in[0,2^K)$ and a vector in $\mathbb{R}^K$ before we explain the generation of the hint. In Table~\ref{kmap}, we show the relation between index $j$, the canonical vector ${\bf e}_j \in \mathbb{R}^K$ and the set $\mathcal{I}$ for $K=3$. The hint for any given observation is generated as follows:
\begin{enumerate}
\item We evaluate the AIC for all $j\in [0,2^K)$ using (\ref{AIC}) and the mapping in Table~\ref{kmap}. Let us call the AIC for the $j$-th index as $AIC_j$.
\item We transform the AIC into the range $[0,1]$ as
\beq
    \nu_j=\frac{\exp\left( -AIC_j / 100 \right)}{\sum_{j^\prime=0}^{2^K-1}\exp\left( -AIC_{j^\prime} / 100 \right)}.
\eeq
\item We generate the hint ${\bf h}$ as
  \beq \label{hint}
    {\bf h}=\sum_{j=0}^{2^K-1} \nu_j {\bf e}_j.
\eeq 
\end{enumerate}
Finally, the indices in  ${\bf h}$ ($+1$ for consistency) that have values $>0.5$ are selected to be part of $\mathcal{I}$ and be part of calibration. Consequently, we design the actor in the SAC algorithm to produce an action in $[0,1]^{K}$, or in other words, the predict the probability of each $i$ being selected to $\mathcal{I}$. In section \ref{results}, we will discuss the RL agent in detail.

\begin{table}[h]
\caption{Mapping between $2^K$ and $K$ for $K=3$} \label{kmap}
\begin{center}
\begin{tabular}{ccc}
  $j$  & ${\bf e}_j$ & $\mathcal{I}$ \\
\hline \\
  0 & $[0\ 0\ 0]^T$ & $\{\ \}$\\
  1 & $[0\ 0\ 1]^T$ & $\{1 \}$\\
  2 & $[0\ 1\ 0]^T$ & $\{2 \}$\\
  3 & $[0\ 1\ 1]^T$ & $\{1, 2 \}$\\
  \vdots & \vdots & \vdots\\
  7 & $[1\ 1\ 1]^T$ & $\{1, 2, 3 \}$\\
\end{tabular}
\end{center}
\end{table}

\section{Results\label{results}}
We present results based on several environments in this section. The focus is to test the sample efficiency of hint assisted RL (expected to be similar to model based RL in sample efficiency) compared with model free RL. The hyperparameters used in all experiments are given in Table~\ref{hyp_all}.

\begin{table*}[h]
\caption{Hyperparameters used in all experiments} \label{hyp_all}
\begin{center}
\begin{tabular}{llll}
  Hyperparameter & Bipedal walker & Elastic net& Calibration\\
\hline \\
  learning rate $\lambda$ & 3e-4 & 1e-3 & 3e-4 \\
  batch size & 256 & 64 & 256\\
  $\tau$ & 0.005 & 0.005& 0.005\\
  $\alpha$ & 0.036& 0.03 & 0.04\\
  $\gamma$ & 0.99 & 0.99 & 0.99 \\
  reward scale &$\times 5$ if $>0$ & $\times 2$ & $\times 10$ if $>0$\\
  $\rho$ & 0.001 & 0.01 & 1.0\\
  constraint $c(\cdot,\cdot)<\delta$& MSE& MSE & KLD\\
  & $\delta=0.5$& $\delta=0.1$ & $\delta=0.01$\\
  $C$ & 10 & 10 & 10 \\
\end{tabular}
\end{center}
\end{table*}

\subsection{Bipedal walker}
\begin{figure}[ht]
\begin{minipage}{0.98\linewidth}
\begin{center}
\epsfig{figure=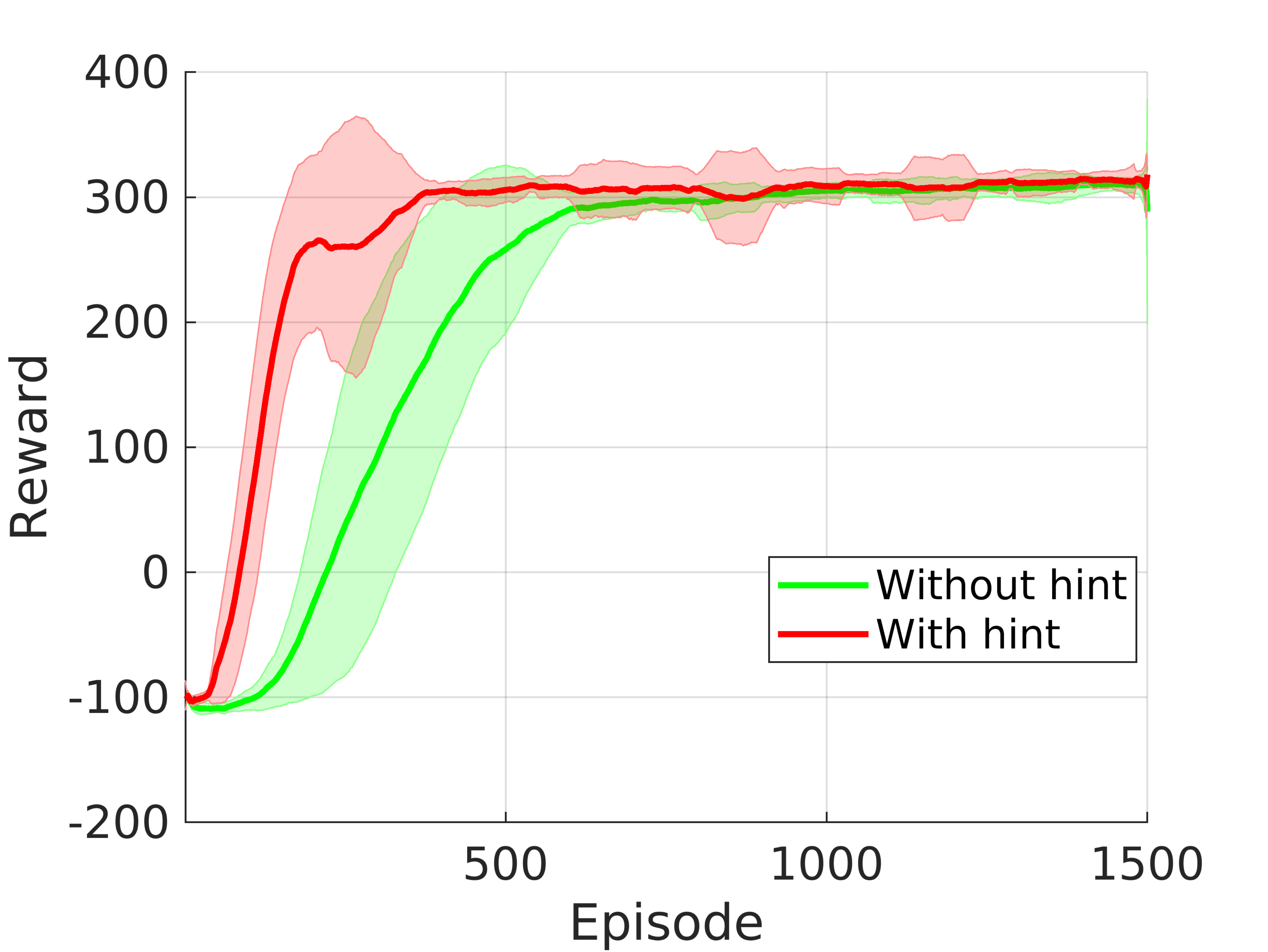,width=8.0cm}\\
\end{center}
\end{minipage}
  \caption{The performance of SAC in the {\tt BipedalWalker-v3} environment. The agent assisted by hints learns faster.\label{bipedal}}
\end{figure}

We consider the {\tt BipedalWalker-v3} and {\tt BipedalWalkerHardcore-v3} environments provided by {\em openai.gym}. In all experiments with this environment, we use the standard two layer network model for the actor and the critic. We train the agent for one realization of the {\tt BipedalWalker-v3} until it converges and we save the models for the actor and the critic to provide hints in all experiments that follow. In Fig. \ref{bipedal}, we show the performance of agents over $10$ runs with different random seed initialization. Note that the hints are provided by a separate agent initialized with the saved model that is reused in all runs.

\begin{figure}[ht]
\begin{minipage}{0.98\linewidth}
\begin{center}
\epsfig{figure=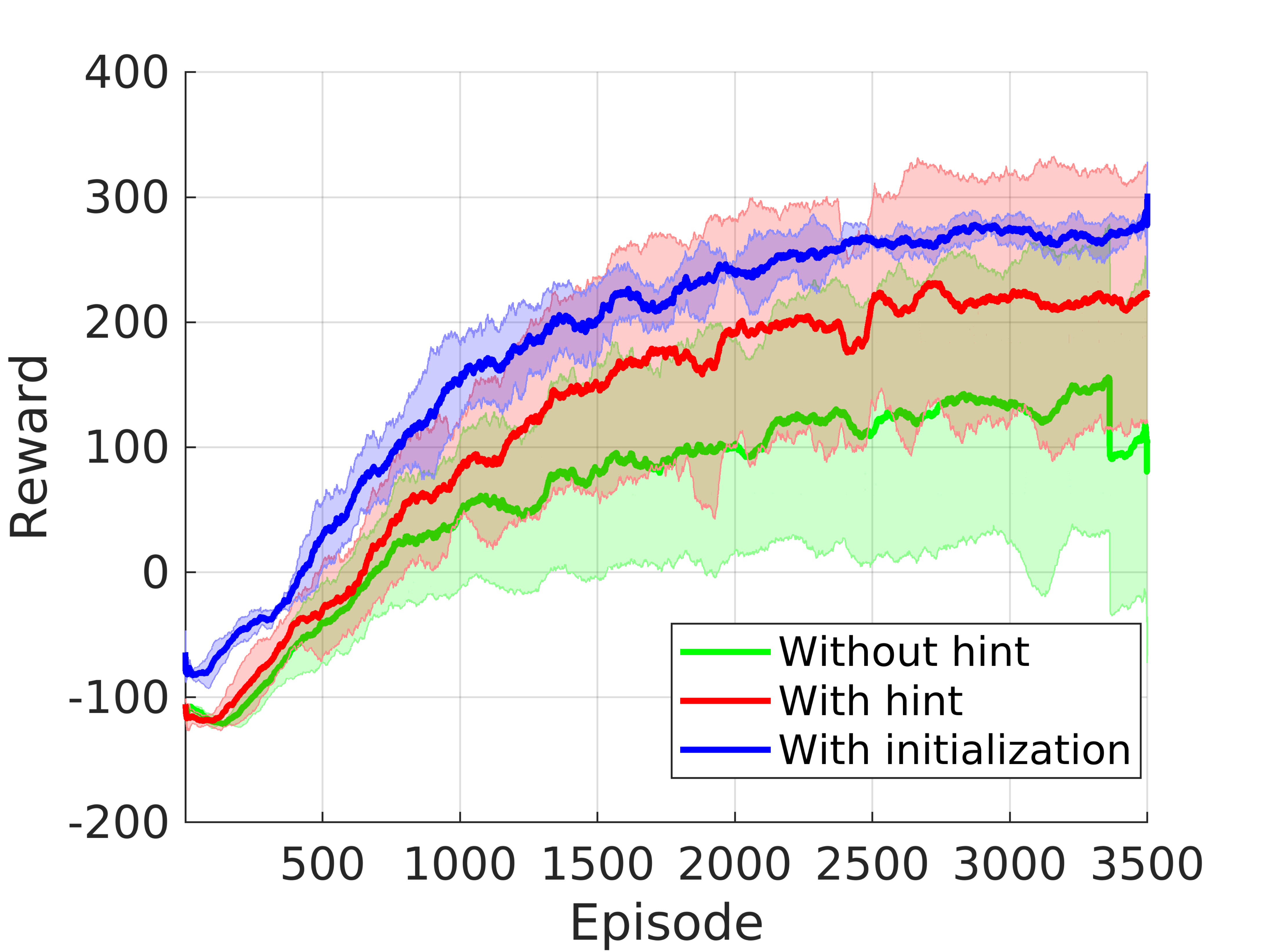,width=8.0cm}\\
\end{center}
\end{minipage}
  \caption{The performance of SAC in the {\tt BipedalWalkerHardcore-v3} environment. The agent assisted by hints learns faster, but not as fast as an agent that is initialized with the model that is being used to generate the hints.\label{bipedal_hardcore}}
\end{figure}

We see in Fig. \ref{bipedal} that by using hints, we can learn faster compared to having no hints. In the next experiment, we still keep the saved model to provide hints (learnt from {\tt BipedalWalker-v3}) and train the agent to solve the {\tt BipedalWalkerHardcore-v3} environment. Note that with this setting, the hints provided are  to a general extent inaccurate. To accommodate this inaccuracy, we tune the values of $\rho$ and $\delta$ in Algorithm \ref{algSAC} as shown in Table~\ref{hyp_all}. In Fig. \ref{bipedal_hardcore}, we show the results in {\tt BipedalWalkerHardcore-v3} environment averaged over 4 runs with different random seed. We also show the result for an agent who is initialized by the saved model that was used to generate the hints. Obviously this means that the networks used for both experiments {\tt BipedalWalker-v3} and {\tt BipedalWalkerHardcore-v3} are the same. Using hints shows a clear improvement as opposed to using no hints. However, compared to initialization of the model with the saved model, using hints needs more iterations to reach the highest expected reward.

\subsection{Elastic net regression}
Given the observation ${\bf x}$ ($\in \mathbb{R}^M$) and the linear model ${\bf x}={\bf A}{\bmath \theta}$ with design matrix ${\bf A}$ ($\in \mathbb{R}^{M\times M}$), elastic net regression \citep{ElasticNet} finds a solution for ${\bmath \theta}$ as
\beq
\widehat{{\bmath \theta}}=\underset{\bmath \theta}{\argmin}\left(\|{\bf x}-{\bf A}{\bmath \theta}\|^2 + \rho_2 \|{\bmath \theta}\|^2+ \rho_1 \|{\bmath \theta}\|_1 \right).
\eeq
Model free RL has been used to determine the best values for $\rho_1$ and $\rho_2$ by \cite{Y2021}. In this experiment, we compare the model free RL performance with the performance using hints to determine the best values for $\rho_1$ and $\rho_2$. We use grid search to generate a hint for $\rho_1$ and $\rho_2$, albeit on a coarse grid.

The state vector is a concatenation of $\mathrm{vec}({\bf A})$ and the eigenvalues $(1+\lambda(\mathcal{H}))$ where 
\beq
\mathcal{H} = {\bf A}\frac{1}{2}\left({\bf A}^T{\bf A}+(\rho_2+\rho_1 \delta(\|{\bmath \theta}\|)) {\bf I}\right)^{-1}\left(-2{\bf A}^T\right).
\eeq
The reward is evaluated as $\frac{\|{\bf x}\|}{\|{\bf x}-{\bf A}{\bmath \theta}\|}+\frac{min(1+\lambda(\mathcal{H}))}{max(1+\lambda(\mathcal{H}))}$.

\begin{figure}[ht]
\begin{minipage}{0.98\linewidth}
\begin{center}
\epsfig{figure=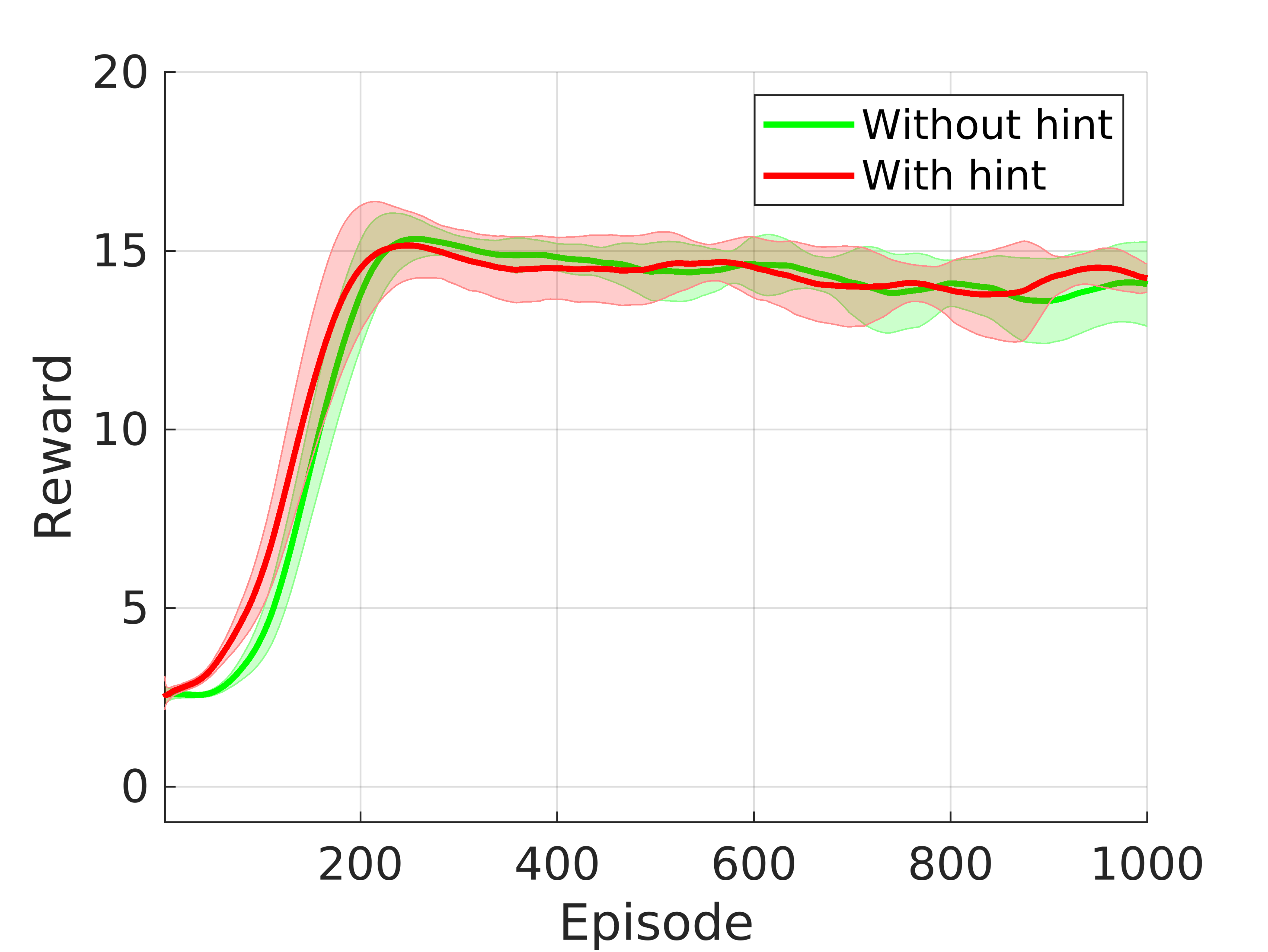,width=8.0cm}\\
\end{center}
\end{minipage}
  \caption{The performance of SAC in solving the elastic net regression hyperparameter selection.\label{enet}}
\end{figure}

In Fig. \ref{enet} we show the performance of SAC in solving the elastic net environment (20 different realizations) for a problem size of $M=20$. The number of steps per episode is limited to 10. Using hints shows only a slight improvement initially but both methods reach the solution almost at the same rate. The lack of a major difference in Fig. \ref{enet} we attribute to the simplicity of the problem.

\subsection{Radio interferometric calibration}
We use a specialized simulator to simulate observations of the LOFAR radio telescope \citep{LOFAR} in training the RL agent. We consider $K=5$ well known sources as outliers in the sky (namely Cassiopeia A, Cygnus A, Taurus A, Virgo A and Hercules A). We vary the target direction, observing frequency, observing epoch, receiver noise and systematic errors in each simulation. We also make sure each observation is valid, i.e., the target remains above the horizon.

The state is a concatenation of the influence map \citep{Y2021} of each calibration (an image of 128$\times$128 pixels), the angular separation of each $K$ sources from the target and some metadata including the azimuth and elevation of each source, observing frequency and $N$. The action is the $K$ probabilities of each source being selected into $\mathcal{I}$ (the output of the actor is in $[-1,1]^K$ which is transformed into $[0,1]^K$). The reward is calculated using the negative AIC given in (\ref{AIC}). We find the AIC when $\mathcal{I}=\{\ \}$, i.e., when only the target is selected for calibration and subtract it from the reward. We also scale the reward to have a standard deviation of about 1.

\begin{figure}[ht]
\begin{minipage}{0.98\linewidth}
\begin{center}
\epsfig{figure=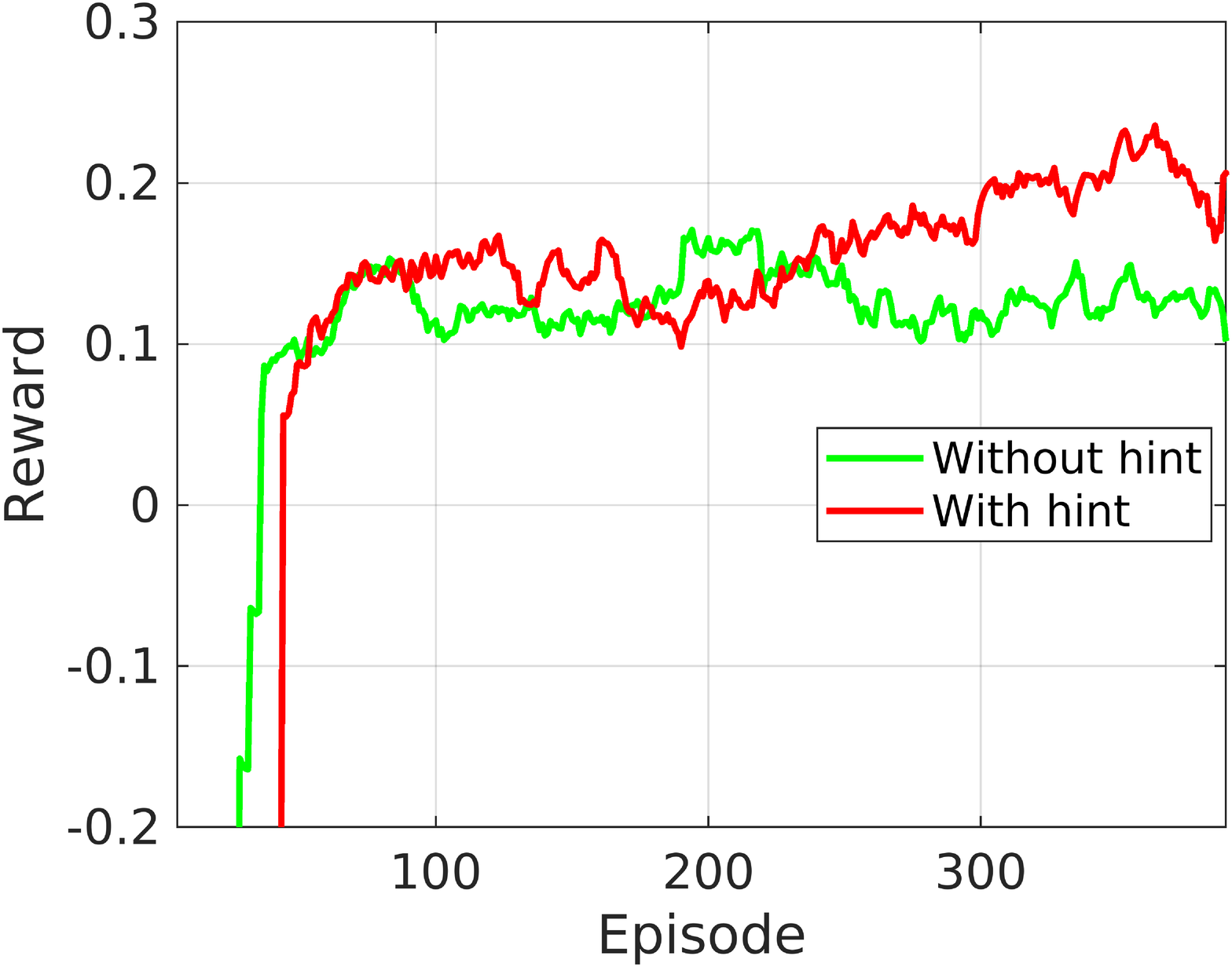,width=8.0cm}\\
\end{center}
\end{minipage}
  \caption{The performance of SAC in solving selection of $\mathcal{I}$ in calibration.\label{demix}}
\end{figure}

In Fig. \ref{demix}, we show the reward obtained by SAC agent with and without hints. The number of steps per each episode (an episode is one simulated observation) is kept at 7. As seen in Fig. \ref{demix}, using hints enable us to achieve a slightly higher reward. We also have seen that the actor is prone to get stuck in local minima, producing the same action for all observations. This is significantly reduced by using hints. Due to the same reason, we have used large value for $\rho$ and a small value for $\delta$ as given in Table~\ref{hyp_all}. 

\begin{figure}[ht]
\begin{minipage}{0.98\linewidth}
\begin{center}
\epsfig{figure=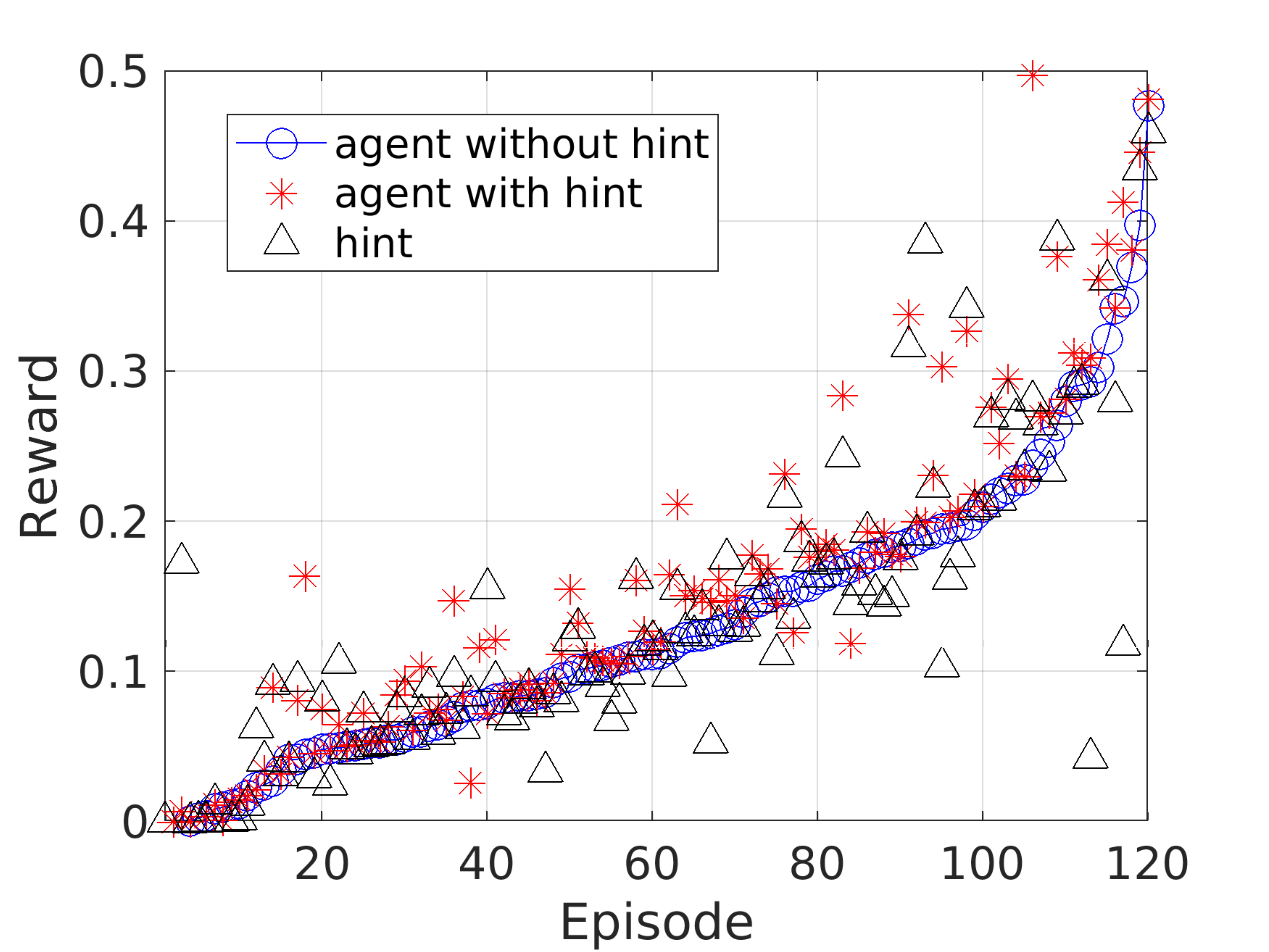,width=8.0cm}\\
\end{center}
\end{minipage}
  \caption{Performance evaluation of the trained agents (after about 4000 training episodes). The evaluation episodes are ordered according to the reward gained by the agent trained without using hints. The agent trained with hints gives a better or equal reward in almost all episodes. Using the hint itself as the action gives mixed results, some being better and some being worse.\label{demix_episodes}}
\end{figure}

In Fig. \ref{demix_episodes} we show the rewards obtained by agents that are trained both with and without using hints. The training used about 4000 episodes. In addition, we also show in Fig. \ref{demix_episodes} the reward obtained by just using the hint as the action. In each randomly generated episode, the agents take $7$ steps and the action giving the maximum reward is taken to be the solution. We order the episodes by the sorted reward obtained by the agent trained without hints, for clarity. It is clear from Fig. \ref{demix_episodes} that in most episodes, the agent trained using hints perform better than (or equal to) the agent trained without hints. The converse is happening in only a handful of episodes. It is also noteworthy to see that just using the hint as the action gives mixed results, being both better and worse than the trained agents.

\section{Conclusions \label{conc}}
When an alternative method exists to generate actions to take in any given state, we have modified the SAC algorithm to use such actions as hints in learning the policy. Results based on several environments show that it is beneficial to use hints, even when the hints are not entirely accurate. Further investigations can be done in using stochastic hints as opposed to deterministic hints as done in this work. The source code for the software is available online \footnote{https://github.com/SarodYatawatta/hintRL}.

\section*{Acknowledgements}
We thank the anonymous reviewers for the careful review and helpful comments.

\bibliographystyle{apalike}
\bibliography{references}

\end{document}